\documentclass[twocol]{ametsoc}

\journal{jas}

\bibpunct{(}{)}{;}{a}{}{,}

\newcommand{\avg}[1]{\left\langle#1\right\rangle}

\newcommand{\pp}[2]{\frac{\partial#1}{\partial #2}}
\newcommand{\partialx}[1]{\frac{\partial#1}{\partial x}}
\newcommand{\partialy}[1]{\frac{\partial#1}{\partial y}}
\newcommand{\partialq}[1]{\frac{\partial#1}{\partial q'}}
\newcommand{\partialt}[1]{\frac{\partial#1}{\partial t}}
\newcommand{\qbar}{\overline q}
\newcommand{\qmax}{q_{\rm max}}
\newcommand{\qmin}{q_{\rm min}}
\newcommand{\rhavg}{r_{\rm bin}}
\newcommand{\rhpara}{r_{\rm para}}

\newcommand{\dd}{{\rm d}}
\newcommand{\Eq}[1]{(\ref{#1})}
\newcommand{\Eqs}[2]{(\ref{#1}) and (\ref{#2})}

\newcommand{\FF}[1]{Figure\,\ref{#1}}
\newcommand{\Fig}[1]{Fig.\,\ref{#1}}
\newcommand{\Heaviside}{\mathcal H}
\newcommand{\sect}[1]{Section\,\ref{#1}}
\newcommand{\sects}[2]{Sections\,\ref{#1} and \ref{#2}}
\newcommand{\tPhio}{\tilde\Phi_*}
\newcommand{\Phio}{\Phi_*}
\newcommand{\Ceff}{C_{\rm eff}}
\usepackage{microtype}
\usepackage{newtxtext}
\usepackage{newtxmath}
\usepackage{bm}
\usepackage{balance}
\renewcommand{\vec}{\bm}

\title{A Stochastic Lagrangian Basis for a Probabilistic Parameterization of Moisture Condensation in Eulerian Models}

\authors{Yue-Kin Tsang
\correspondingauthor{Yue-Kin Tsang, School of Mathematics, University of Leeds, Leeds, LS2 9JT, United Kingdom.}}

\affiliation{School of Mathematics, University of Leeds, Leeds, United Kingdom}

\email{y.tsang@leeds.ac.uk}

\extraauthor{Geoffrey K. Vallis}

\extraaffil{College of Engineering, Mathematics and Physical Sciences, University of Exeter, Exeter, United Kingdom } 

\abstract{In this paper we describe the construction of an efficient probabilistic parameterization that could be used in a coarse-resolution numerical model in which the variation of moisture is not properly resolved. An Eulerian model using a coarse-grained field on a grid cannot properly resolve regions of saturation---in which condensation occurs---that are smaller than the grid boxes. Thus, in the absence of a parameterization scheme, either the grid box must become saturated or condensation will be underestimated. On the other hand, in a stochastic Lagrangian model of moisture transport, trajectories of parcels tagged with humidity variables are tracked and small-scale moisture variability can be retained; however, explicitly implementing such a scheme in a global model would be computationally prohibitive. One way to introduce subgrid-scale saturation into an Eulerian model is to assume the humidity within a grid box has a probability distribution. To close the problem, this distribution is conventionally determined by relating the required subgrid-scale properties of the flow to the grid-scale properties using a turbulence closure. Here, instead, we determine an assumed probability distribution by using the statistical moments from a stochastic Lagrangian version of the system. The stochastic system is governed by a Fokker--Planck equation and we use that, rather than explicitly following the moisture parcels, to determine the parameters of the assumed distribution. We are thus able to parameterize subgrid-scale condensation in an Eulerian model in a computationally efficient and theoretically well-founded way. In two idealized advection--condensation problems we show that a coarse Eulerian model with the subgrid parameterization is well able to mimic its Lagrangian counterpart.}

\begin{document}

\maketitle

\section{Introduction}
\label{intro}

Water vapor is carried around as a tracer, normally condensing when the vapor pressure reaches the saturation value given by the Clausius--Clapeyron relation. (Strictly, the Clausius--Clapeyron relation determines the vapor pressure for water, or other condensible, in thermodynamic equilibrium with liquid water or ice, and if neither are present then the water vapor can become supersaturated, but for the purposes of this paper we will assume condensation occurs upon saturation.) Furthermore, condensation normally occurs quickly upon saturation, and it is common in the atmospheric sciences to refer to the `fast condensation limit'. The liquid water produced by condensation may then, in simple models, be assumed to fall to the ground, or, more realistically it may be advected by the flow before the drops coalesce and fall as rain or form ice, as determined in models by more or less complicated microphysical parameterization schemes.

Putting aside the complications of microphysical effects, the simplest advection--condensation model consistent with this picture postulates that as an air parcel is advected by the large-scale wind, and away from evaporation sources, the moisture content of the parcel remains constant except when it exceeds the local saturation limit, at which point the excessive water vapor is removed by condensation. Previous work, e.g. \cite{Salathe97, Pierrehumbert98, Galewsky05, Dessler07}, has applied this idea to reconstruct large-scale features of the atmospheric moisture distribution. In these studies, the trajectory of a parcel is traced backward in time to the location where the parcel is last saturated, e.g. the point at which it last encounters convection or the lower boundary layer. The specific humidity of the parcel at its present location is then given by the minimum saturation specific humidity encountered along the trajectory. The success of these studies highlights the Lagrangian nature of the large-scale transport and condensation of atmospheric moisture. Indeed, \citet{Pierrehumbert07} suggested that the proper approach to represent moisture transport in climate models is to take the stochastic Lagrangian viewpoint whereby the fluctuations in the trajectories of moist parcels are parameterized by random processes. 

Over the past few decades, stochastic Lagrangian models, which describe the trajectories of air parcels using a model of random velocity, have been developed to study turbulent transport in the atmosphere \citep{Wilson96,Thomson13}. Under a Markov assumption, the parcel position and velocity are random variables satisfying some stochastic differential equations. Equivalently, the model can also be specified by a Fokker--Planck equation which governs the joint probability density function (PDF) of position and velocity. Applying this approach to moisture transport, various studies \citep{OGorman06, Pierrehumbert07, Sukhatme11, Beucler16, Tsang17} have investigated theoretically the advection--condensation of water vapor by evolving an ensemble of particles, each carrying its own set of dynamical and thermodynamical variables obeying stochastic model equations. Whereas the stochastic Lagrangian description does have the advantage of, in principle, retaining local fluctuations at small scales, it also comes with a high computational cost---it is simply impractical to carry around a very large number of Lagrangian particles representing moist air parcels. One possible way to address this problem is to use a hybrid parcel-in-cell method \citep{Dritschel18}, but below we will describe a qualitatively different approach, in which the resulting equations are Eulerian (and so can be efficiently solved) but the underlying parameterization is explicitly based on a Lagrangian description.

The conventional practice is to represent atmospheric water vapor as a coarse-grained field on a numerical grid, writing the equations of motion in the Eulerian form as a partial differential equation (PDE), for example
\begin{equation}
	\label{eq:eulm1} 
   	\pp q t + \vec u \cdot \nabla q = \nabla\cdot(D \nabla q) + S - C.
\end{equation}
In this equation, $\vec u$ is velocity, $q$ is specific humidity, $S$ is a moisture source, $C$ represents the effects of condensation and $D(\vec x,t)$ is a diffusivity. The condensation term is zero until saturation occurs. In reality, $D$ would be the molecular diffusivity and is very small indeed, so that the specific humidity of an unsaturated parcel is essentially conserved. However, in a coarse-resolution model---such as a climate model with a horizontal resolution measured in kilometers---$D$ is often a parameterized diffusivity much larger than the molecular one. It also cannot be small for numerical reasons. (A semi-implicit, semi-Lagrangian scheme may be stable at low resolution without a high explicit diffusivity, but these methods can also be diffusive or inaccurate.) Furthermore, if condensation is only allowed to occur at saturation, then the effects of diffusion are in many circumstances such as to make large regions saturated, producing moisture fields noticeably different from a Lagrangian model \citep[][Chapter 18]{Pierrehumbert07,  Vallis17}. In a climate model with a moisture equation similar to \Eq{eq:eulm1}, condensation and rainfall will only occur when a grid box is entirely saturated. This has long been recognized to be in many circumstances quite unrealistic \citep[e.g.,][]{Sommeria77} and, because of the strong dependence of the absorption of outgoing long-wave radiation on water content, such a misrepresentation can be especially significant in the modeling of the Earth's radiation budget.

The problem with the Eulerian approach, as noted by \cite{Pierrehumbert07}, is that the coarse-graining that is in practice required does not commute with the highly nonlinear condensation process. Our first goal is in fact to demonstrate theoretically how this causes an Eulerian model without subgrid-scale condensation to produce large regions of saturation compared to its Lagrangian counterpart. A possible solution to such problems, sometimes used in cloud modeling \citep{Tompkins02, Jakob02}, is to suppose that the specific humidity (and possibly other thermodynamical variables) inside a given grid box is not single-valued but has a probability distribution, thus introducing local fluctuations into the system. Then, part of the box may be saturated even though the average specific humidity over the box is less than the saturation limit, and some fraction of the water vapor content, as determined by the probability function, may then be removed by condensation. A probabilistic parameterization of subgrid-scale condensation along these lines was proposed by \citet{Sommeria77} and \citet{Mellor77} to model moist convection in the boundary layer. They assumed the total mixing ratio and the liquid potential temperature have a joint Gaussian PDF and determined the cloud fraction within a grid cell from such a PDF. \citet{Bougeault82} later used this `assumed PDF' method with several different PDF shapes to model the trade-wind cumulus layer. Since then numerous variations have been developed and employed in atmospheric numerical models. For example, one of the cloud schemes in the Met Office Unified Model is the \cite{Smith90} scheme which uses a triangular PDF \citep{Wilson08}, a somewhat more complicated scheme has been used at ECMWF \citep{Tiedtke93}, and various other, sometimes still more complicated (and computationally intensive) schemes have been proposed \citep[e.g.,][]{Lappen01, Tompkins02, Golaz02, Yoshida10, Bogenschutz13}. A crucial step in these probabilistic schemes is to determine the parameters of the prescribed PDF, such as width and skewness. This is often done by linking the PDF parameters to various eddy fluxes or correlation functions, and turbulence closure models are then used to predict these correlations from the resolved scales. The difficulty with these approaches is, of course, that the parameterization is only as good as the turbulence closure it is based upon. 

Evidently, then, both Lagrangian and Eulerian approaches have advantages and shortcomings---the former is accurate but impractical, the latter is practical but less accurate, with the contrast stemming from the fundamental differences in the representation of particle motion and condensation of the two formulations. In this paper we seek to combine these two approaches. The idea is to use information extracted from a corresponding stochastic Lagrangian model to derive a parameterized Eulerian model that can produce similar results to the stochastic Lagrangian model, but at a fraction of the computational cost. We aim to achieve two goals. The first is to provide a sound theoretical basis to the heuristic probabilistic schemes that are in common use. The second is to describe a systematic way whereby a probabilistic parameterization for the condensation in an Eulerian model may be derived, for example to provide a $C$ in the Eulerian equation \eqref{eq:eulm1}.

The premise of our method is that the small-scale velocity $\vec V'$ of a moist parcel can be modeled as a random process. We may then represent the moist dynamics by a stochastic Lagrangian model in which an ensemble of moist parcels is advected by the velocity $\vec V + \vec V'$ where $\vec V$ is the deterministic large-scale parcel velocity. Since $\vec V'$ is random, at each location and time the stochastic system produces a PDF of the humidity, $\hat P(q|\vec x,t)$. Now, it is expensive to obtain $\hat P$ by performing Monte Carlo simulation of the stochastic differential equations or by solving the high dimensional Fokker--Planck equation governing $\hat P$. Instead, we use an assumed PDF $\Phio$ as surrogate for $\hat P$ and require the moments of $\Phio$ to match those of $\hat P$ derived from the Fokker--Planck equation. The $\Phio$ so determined will then be used in a probabilistic parameterization of $C$ for an Eulerian model such as \Eq{eq:eulm1}. Thus our scheme involves two steps. First, an appropriate stochastic Lagrangian model must be constructed, and second the Fokker--Planck equation---as an alternative to turbulence closures---must be used to derive parameters for an assumed PDF. We carry out this procedure in two idealized advection--condensation problems and show that a coarse Eulerian model with the subgrid parameterization is, in fact, well able to mimic its Lagrangian counterpart.  Because of the idealized nature of these problems, we are able to solve the Lagrangian model directly, by Monte Carlo simulations of moist particles advected by a large-scale field and a random component, and so provide a true test of the methodology.

The paper is organized as follows. In \sect{ac} we present the basic Lagrangian and Eulerian methodologies using a model of moisture transport in an overturning cell, and show that an Eulerian model tends to produce saturated air. \sect{paracond} gives the details of probabilistic parameterization of condensation. We then compare results from Eulerian models with and without parameterization to those of Lagrangian models for a steady flow in \sect{cell} and for an unsteady flow in \sect{wave}. In \sect{sec:slmpara} we discuss the use of an underlying stochastic Lagrangian model to parameterize condensation in coarse-grained atmospheric models, and we conclude the paper in \sect{sec:conclude}.

\section{Lagrangian particles versus Eulerian fields}
\label{ac}

\subsection{Advection--condensation in an overturning cell}
\label{ac_cell}

We consider the advection of moist air in a square domain $[0,\pi] \times [0,\pi]$ on the $xy$-plane. Condensation occurs as water vapor is transported by a prescribed velocity through a saturation specific humidity field $q_s$. We assume the velocity has an incompressible large-scale component $\vec u = (u,v)$ and a turbulent component at the small scales. In this section, as a crude model with some similarities to the Hadley cell, we take $(u,v)=(-\partial_y\psi,\partial_x\psi)$ as a steady overturning flow with streamfunction
\begin{equation}
\psi(x,y) = \sin x \sin y.
\label{psi1}
\end{equation}
\FF{cell_qs} shows the streamlines of $\vec u$ in a schematic of the system. We assume $q_s$ varies only with the altitude $y$ and is independent of time. Specifically, we assume a linear temperature profile in $y$:
\begin{equation}
T(y) = T_{\max} - (T_{\max}-T_{\min})\frac{y}{\pi}.
\end{equation}
Using an empirical Magnus or Tetens formula \citep{Bolton80,Lawrence05} for the saturation vapor pressure,
\begin{equation}
e_s(T) = 6.112\exp\left(\frac{17.67\,T}{T+243.3}\right) \text{hPa},
\end{equation}
together with $q_s\approx 0.622 e_s/(1010\,\text{hPa})$ gives
\begin{equation}
q_s(y) = 3.619\times 10^{-3} \exp\left[\frac{17.67\,T(y)}{T(y)+243.3}\right]
\label{qsat}
\end{equation}
and we define
\begin{equation}
\qmin\equiv q_s(\pi), \quad \qmax\equiv q_s(0).
\end{equation}
Here, we set $T_{\max}=26\,^{\circ}$C and $T_{\min}=-50\,^{\circ}$C. Hence, $\qmax=0.019$ and $\qmin=3.7\times 10^{-5}$. We assume there is an evaporation source $S$ located at the bottom boundary to maintain the specific humidity along $y=0$ at $\qmax$.

\begin{figure}
\centering
\includegraphics[width=0.47\textwidth]{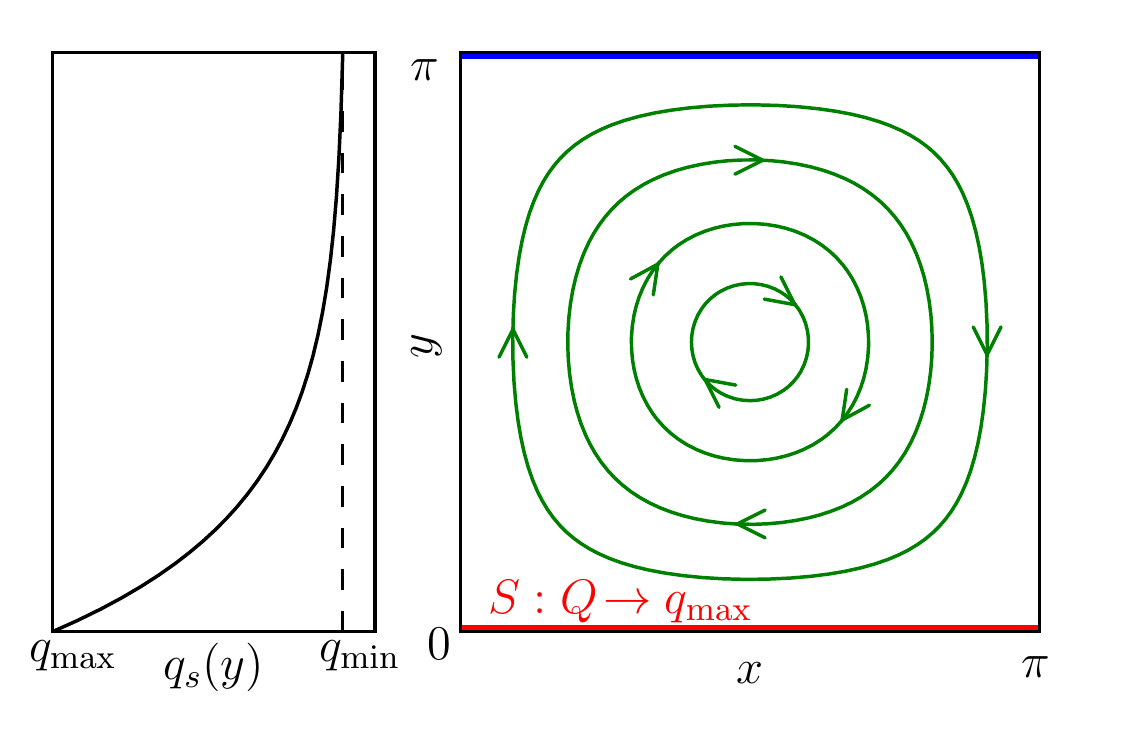}
\caption{Schematic of moisture transport in an overturning cell described in \sect{ac}\ref{ac_cell}. The streamlines of the large-scale circulation \Eq{psi1} is shown as solid lines with arrows and $q_s(y)$ is the saturation specific humidity. An evaporation source $S$ is located at $y=0$ to maintain the moisture in the system.}
\label{cell_qs}
\end{figure}

The interplay between large-scale coherent flow, small-scale turbulence and condensation in this model roughly reproduces several interesting features of the atmosphere \citep{Tsang17}: a humid bottom boundary layer that resembles the planetary boundary layer, a narrow region of intense condensation along $x=0$ reminiscent of the Tropics and a relative humidity minimum at the center of the cell.

\subsection{Deterministic coarse-grained field formulation}
\label{mf}

For a deterministic Eulerian formulation of the advection--condensation problem described above, the specific humidity is represented by a coarse-grained field $q(x,y,t)$ whose time evolution is governed by the PDE:
\begin{equation}
\partialt{q} + \vec u \cdot\nabla q = \kappa_q \nabla^2 q - C.
\label{pde}
\end{equation}
Above, $\vec u$ is the large-scale velocity and unresolved small-scale turbulence is represented by the diffusion term with constant eddy diffusivity $\kappa_q$. In accord with the advection--condensation paradigm, molecular diffusion is assumed to be negligible. The condensation $C$ may be written as
\begin{equation}
C = \frac{1}{\tau_c} (q-q_s)\,\Heaviside(q-q_s)
\label{cond_euler}
\end{equation}
where $\tau_c$ is the condensation time-scale and $\Heaviside$ is the Heaviside step function. For most of this paper, we employ the rapid-condensation limit of $\tau_c \rightarrow 0$ and implement $C$ as a rule to prevent supersaturation:
\begin{equation}
C: q(x,y,t) \rightarrow \min[\,q(x,y,t)\,,\,q_s(y)\,].
\label{cond_pde}
\end{equation}
The source $S$ is implemented as a boundary condition:
\begin{equation}
q(x,0,t) = \qmax.
\end{equation}
At the other boundaries, we have the no-flux conditions:
\begin{equation}
\left.\partialx{q}\right|_{x=0} = \left.\partialx{q}\right|_{x=\pi}
= \left.\partialy{q}\right|_{y=\pi} = 0.
\end{equation}

\begin{figure*}[t]
\centering
\includegraphics[width=\textwidth]{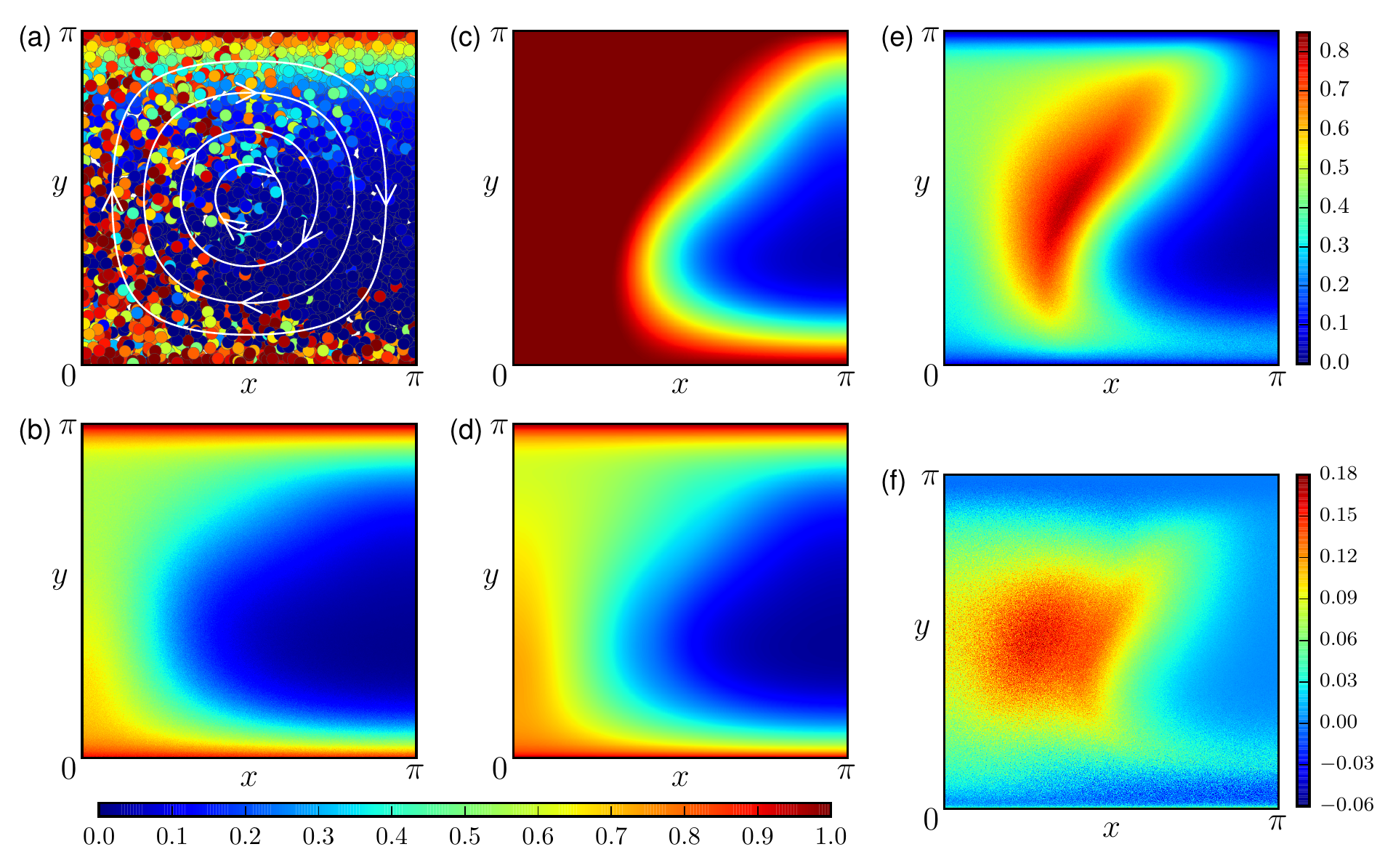}
\caption{Advection--condensation by the overturning flow \Eq{psi1} and with $\kappa=10^{-1}$. (a) Snapshot of the statistically steady state in a Monte Carlo simulation of \Eq{sde}. Color indicates the relative humidity of each parcel. Solid lines are streamlines of \Eq{psi1}. (b) Bin-averaged relative humidity field $\rhavg(x,y)$ calculated from the simulation in (a) as described below \Eq{rhmc}. (c) Steady-state relative humidity field $r(x,y,t)$ at large $t$ from a solution of the Eulerian coarse-grained model \Eq{pde}. (d) Steady-state relative humidity $\rhpara(x,y,t)$ from the same model in (c) but with condensation parameterization implemented as described in \sect{cell}. (e) Deviation of $r(x,y,t)$ in (c) from $\rhavg(x,y)$ in (b). (f) Deviation of $\rhpara(x,y,t)$ in (d) from $\rhavg(x,y)$ in (b).}
\label{mcpde}
\end{figure*}

We solve \Eq{pde} for the field $q(x,y,t)$ using the split-step approach. Given $q(x,y,t_n)$ at time $t_n$, we obtain an intermediate (supersaturated) field $q_*(x,y,t_{n+1})$ by time-stepping forward the advection--diffusion equation
\begin{equation}
\partialt{q} + \vec u \cdot\nabla q = \kappa_q \nabla^2 q
\label{advdif}
\end{equation}
to $t_{n+1} = t_n + \Delta t$.  (Here and elsewhere in the paper we write the equations with partial derivatives with respect to time, such as $\partial q/\partial t$, with the understanding that the procedure takes us from $t_n$ to $t_{n+1}$.)   We use a semi-Lagrangian scheme for the advection and the Alternative Direction Implicit method for the diffusion. We then carry out the condensation \Eq{cond_pde} to produce the moisture field at time $t_{n+1}$:
\begin{equation}
q(x,y,t_{n+1}) = \min[\,q_*(x,y,t_{n+1})\,,q_s(y)\,].
\label{condstar}
\end{equation}
\FF{mcpde}(c) shows the steady-state relative humidity field
\begin{equation}
r(x,y,t) = \frac{q(x,y,t)}{q_s(y)}
\end{equation}
at a large time $t$ from a $513^2$ simulation with $\kappa_q=10^{-1}$.

\subsection{Stochastic particle formulation}

We now turn to a stochastic Lagrangian model of the system. Inside the square domain, the moist air is now represented by an ensemble of air parcels. The domain boundaries are reflective. The parcels are initially uniformly distributed over the domain and will remain so for an incompressible advecting flow.

Let $(X,Y)$ be the position of a parcel and $Q$ be its specific humidity. Consider $(Q,X,Y)$ as random variables, the advection--condensation of each moist parcel is described by the following set of stochastic differential equations:
\begin{subequations}
\begin{align}
\dd X(t) &= u(X,Y) \dd t + \sqrt{2\kappa_b}\,\dd W_1(t), \\
\dd Y(t) &= v(X,Y) \dd t + \sqrt{2\kappa_b}\,\dd W_2(t), \\
\dd Q(t) &= (S-C) \dd t.
\end{align}
\label{sde}%
\end{subequations}
The resolved large-scale velocity $\vec u$ is once again given by \Eq{psi1}.
$W_1(t)$ and $W_2(t)$ are Wiener processes and $\kappa_b$ is the associated Brownian diffusivity. Thus, the small-scale turbulent velocity is modeled as white noise, denoted by $\dot W_i(t)$. In other words, the turbulent velocity of each parcel is a Gaussian random variable at any instance of time $t$ and is uncorrelated in time with correlation function:
\begin{equation}
\overline{\dot W_i(t)\dot W_j(s)}=2\kappa_b\delta(t-s)\delta_{ij},\quad i=1,2.
\end{equation}
Above, $\overline{(\cdot)}$ denotes ensemble average. To match the simulation in the Eulerian formulation, we set $\kappa_b$ equals $\kappa_q$ in anticipation of the discussion surrounding \Eq{qbareqn} and \Eq{pdetau} and denote their common value by $\kappa$:
\begin{equation}
\kappa_b=\kappa_q=\kappa.
\end{equation}
Exchange of moisture between parcels, which may be important in some situations \citep{Haynes97}, is not included in this model and each parcel evolves independently. For finite condensation rate ($\tau_c>0$), $C$ is given by
\begin{equation}
C = \frac{1}{\tau_c} (Q-q_s)\,\Heaviside(Q-q_s)
\label{cond}
\end{equation}
and in the limit of $\tau_c \rightarrow 0$, we have
\begin{equation}
C: Q \rightarrow \min[Q,q_s(Y)].
\label{rpdcond}
\end{equation}
The action of the source $S$ at $y=0$ is that it resets the specific humidity of air parcels to the local saturation value $\qmax$ upon hitting the bottom boundary. For a detailed analysis of this stochastic system, we refer the readers to \cite{Tsang17}.

We perform Monte Carlo simulation of \Eq{sde} using the Euler--Murayama method \citep{Higham01}. At $t=0$, $10^6$ saturated parcels are uniformly distributed over the domain. \FF{mcpde}(a) shows a snapshot of a subset of the parcels after the system has reached a statistically steady state. The color indicates the relative humidity of each parcel
\begin{equation}
R(t) = \frac{Q(t)}{q_s[Y(t)]}.
\label{rhmc}
\end{equation}
To visualize the spatial distribution of moisture over the domain, we construct a bin-averaged field $\rhavg(x,y)$ from the Monte Carlo data by dividing the domain into a uniform gird of square bins. We then average $R(t)$ over all parcels inside the bin centered at $(x,y)$ to obtain $\rhavg(x,y,t)$. As the velocity $\vec u$ is steady and we are interested in the statistically steady distribution, we further average over time to obtain $\rhavg(x,y)$. \FF{mcpde}(b) shows $\rhavg(x,y)$ corresponding to the simulation in \Fig{mcpde}(a), $513^2$ bins have been used. An interpretation of this averaging procedure is that many parcels with different $R(t)$ contribute to a single observation of $\rhavg(x,y,t)$ taken over a small area about $(x,y)$. 

\subsection{Non-commutation between condensation and coarse-graining}

Let us now compare the relative humidity field calculated from the two formulations. As shown clearly in \Fig{mcpde}(b) and \ref{mcpde}(c), the Lagrangian and the Eulerian models produce starkly different results. The Eulerian model has the unrealistic feature that a large part of the domain is fully saturated with $r=1$. \FF{mcpde}(e) plots the difference in the relative humidity field from the two models. Unsurprisingly, the largest discrepancy occurs in the rising half of the cellular flow where most of the condensation happens. Generally, the saturated region in the Eulerian model will shrink as $\kappa_q$ decreases (e.g. compare \Fig{mcpde}(c) and \Fig{kappa2}(a)). However, regardless of the value of $\kappa_q$, the boundary at $x=0$ will remain saturated. This is fundamentally different from the results of the Lagrangian model. \cite{Pierrehumbert07} had observed similar behavior in simple one-dimensional models and attributed it to the loss of local fluctuation in a coarse-grained field representation, and \citet{Vallis17} qualitatively described similar behavior in a two-dimensional model. Here, we investigate this effect quantitatively in the two-dimensional case. 

\begin{figure*}[t]
\centering
\includegraphics[width=0.7\textwidth]{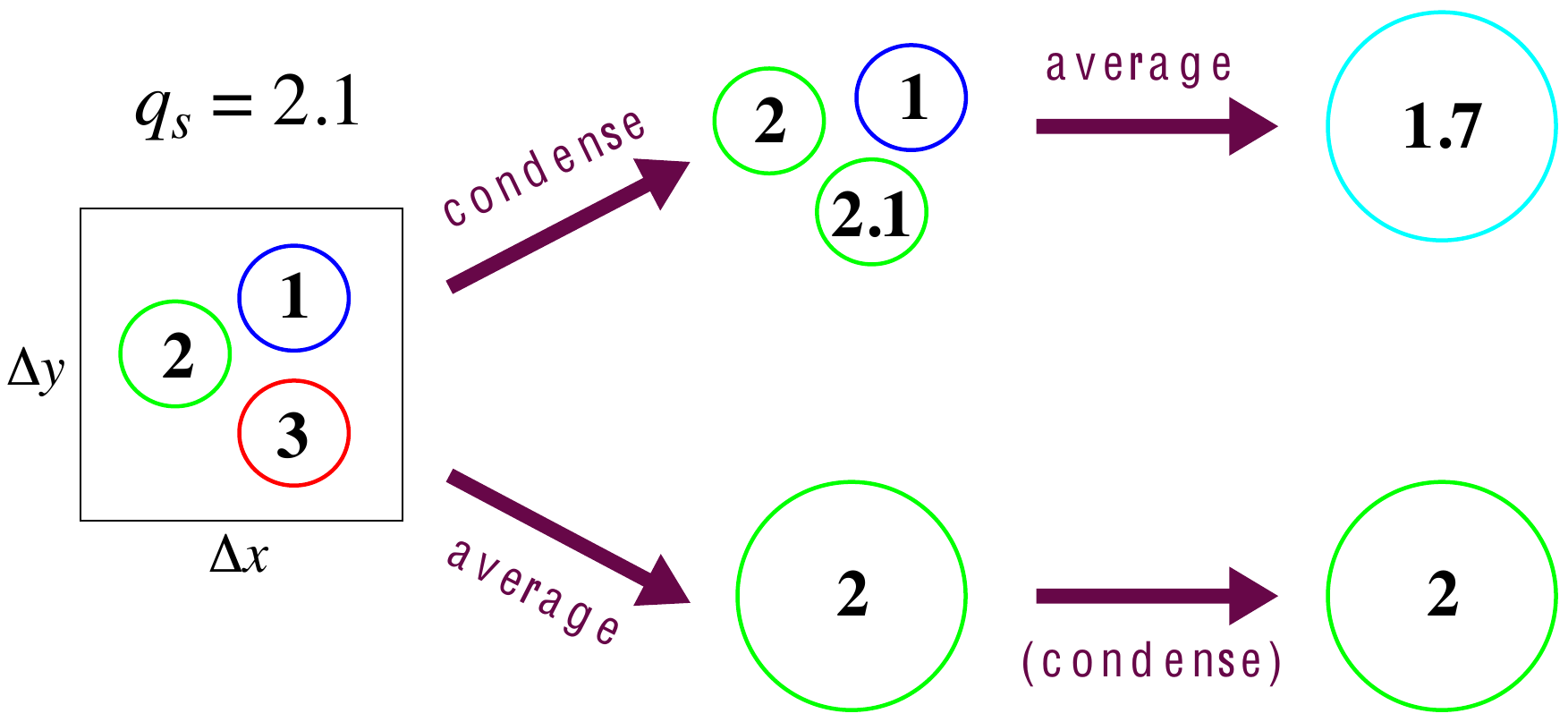}
\caption{Assume there are three moist parcels within an infinitesimal area $\Delta x\Delta y$ where the local saturation value is $q_s$. The numbers inside the circles indicate their specific humidity. In the upper branch, we first condense each parcel individually according to its value of specific humidity and then ``measure" the average value of these condensed parcels. In the lower branch, we first average over the initial specific humidity within $\Delta x\Delta y$ and then carry out the condensation process according to this coarse-grained specific humidity. We see that these two approaches produce different results, with more moisture being retained when averaging precedes condensation. However, note that if all three parcels are initially supersaturated, then the order of condensation and averaging does not matter.}
\label{condfig}
\end{figure*}

\FF{mcpde}(a) clearly shows parcels with a broad range of specific humidity coexist within a small area. When the moisture distribution is represented by a coarse-grained field, such subgrid-scale fluctuation is averaged out leading the system to bias toward saturation. Mathematically, this is because the condensation process and the coarse-graining process do not commute. To elucidate, we examine from a theoretical viewpoint how one goes from \Fig{mcpde}(a) to \Fig{mcpde}(b). To this end, it is more convenient to momentarily revert to a small but non-zero condensation time $\tau_c$. For an ensemble of parcels described by the random variables $(Q,X,Y)$ obeying the stochastic differential equations (\ref{sde}), the joint PDF $P(q',x,y;t)$ of specific humidity and position satisfies the Fokker--Planck equation \citep{Pavliotis14}
\begin{equation}
\partialt{P} + \vec u \cdot\nabla P - \partialq{}(CP) = \kappa_b \nabla^2 P
\label{fptau}
\end{equation}
supplemented by appropriate boundary conditions in the domain $[\qmin,\qmax]\times[0,\pi]\times[0,\pi]$. Above, we have used the incompressibility condition $\nabla\cdot \vec u = 0$ and
\begin{equation}
C(q',y) = \frac{1}{\tau_c} [q'-q_s(y)]\,\Heaviside[q'-q_s(y)].
\label{ctau}
\end{equation}
The mean specific humidity at a given position $(x,y)$ is the conditional expectation value
\begin{equation}
\qbar(x,y,t) = \int_{\qmin}^{\qmax} q' \hat P(q'|x,y;t)\, \dd q'
\label{qbardef}
\end{equation}
where $\hat P(q'|x,y;t)$ is the conditional probability density for a parcel to have specific humidity $q'$ given it is located at $(x,y)$. So the bin-averaged field $\rhavg$ in \Fig{mcpde}(b) is a numerical approximation to $\qbar/q_s$.

We now derive the evolution equation of $\qbar$. By the definition of conditional PDF, $\hat P$ in \Eq{qbardef} is related to $P$ by
\begin{equation}
\hat P(q'|x,y;t) = \frac{P(q',x,y;t)}{p(x,y;t)}
\label{phat}
\end{equation}
where
\begin{equation}
p(x,y;t) = \int_{\qmin}^{\qmax} P(q',x,y;t)\,\dd q'
\label{pmar}
\end{equation}
is the marginal PDF which gives the probability that a parcel is located at $(x,y)$ regardless of its specific humidity. Integrating \Eq{fptau} over $q'$ yields
\begin{equation}
\partialt{p} + \vec u \cdot\nabla p  = \kappa_b \nabla^2 p.
\label{marginal}
\end{equation}
Note that the boundary term involving $C$ from the integration vanishes. This is because $C(\qmin,y)=0$ by \Eq{ctau} and also as $\tau_c\rightarrow 0$, no parcel can have $Q=\qmax$ inside the domain: $P(\qmax,x,y;t)\rightarrow 0$ for all $y>0$. It then follows from \Eq{fptau} and \Eq{marginal} that $\hat P$ satisfies
\begin{equation}
\pp{\hat P}{t} + \left(\vec u - \frac{2\kappa_b}{p}\nabla p\right)\cdot\nabla\hat P - \pp{}{q'}(C\hat P) = \kappa_b \nabla^2 \hat P \,.
\label{phateq}
\end{equation}
Since the parcels are uniformly distributed at $t=0$ in our setup, \Eq{marginal} implies $p=\pi^{-2}$ for all $t$, hence the parcels remain uniformly distributed and the term involving $\nabla p$ in \Eq{phateq} vanishes. This simply means we are concerned with a constant air density.

Multiply \Eq{phateq} by $q'$ and integrate over $q'$, we finally get the equation for $\qbar$:
\begin{equation}
\partialt{\qbar} + \vec u\cdot\nabla\qbar = \kappa_b \nabla^2 \qbar
-\frac{1}{\tau_c}\int_{\qmin}^{\qmax}(q'-q_s)\Heaviside(q'-q_s)\hat P\,\dd q'.
\label{qbareqn}
\end{equation}
On the other hand, the governing equation (\ref{pde}) of $q(x,y,t)$ in the Eulerian formulation for non-zero $\tau_c$ reads:
\begin{equation}
\partialt{q} + \vec u\cdot\nabla q  = \kappa_q \nabla^2 q - \frac{1}{\tau_c}(q-q_s)\Heaviside(q-q_s).
\label{pdetau}
\end{equation}
Comparing \Eq{qbareqn} with \Eq{pdetau}, we see that the differences in $\qbar$ and $q$ stem from the condensation term. In \Eq{qbareqn}, condensation for each individual parcel is considered before their contributions to $\qbar$ are added up. Thus, local fluctuations are accounted for. In \eqref{pdetau}, only the coarse-grained value $q$ is available and condensation only happens when $q>q_s$, causing the system to retain more moisture as seen in \Fig{mcpde}(c). \FF{condfig} illustrates this non-commutation between condensation and coarse-graining pictorially with an example.

\section{Probabilistic parameterization of condensation}
\label{paracond}

\subsection{An effective condensation}

We have seen in previous sections that modeling water vapor distribution using a coarse-grained field is prone to producing saturation. On the other hand, the Lagrangian approach is able to produce more realistic results, albeit with higher computational cost, by accounting for the effects of subgrid-scale moisture fluctuation on condensation. Here we ask the question: if we regard the Lagrangian model as `truth', how do we construct an Eulerian PDE-based model that might be used in its place to give similar results? Comparing \Eqs{qbareqn}{pdetau} suggests naturally the answer is to replace the condensation term in the Eulerian equation (\ref{pdetau}) by an effective condensation
\begin{equation}
\Ceff = \frac{1}{\tau_c}\int_{q_s(y)}^{\qmax}(q'-q_s)\Phio(q'|x,y;t)\,\dd q'
\label{Ceff}
\end{equation}
where $\Phio(q'|x,y;t)$ is an approximation to the `true' conditional PDF $\hat P(q'|x,y)$ in the Lagrangian model. Equation (\ref{Ceff}) resembles the formula for liquid water content in a conventional probabilistic subgrid-scale cloud scheme \citep{Sommeria77}. To specify $\Phio$, we take the `assumed PDF' approach by assuming a functional form for $\Phio$ that contains a small number of parameters. These parameters are then determined by matching the moments of $\Phio$ to those of $\hat P$ up to a certain order. Note that $\Phio$ is not governed by an evolution equation and there is the freedom to assume different functional forms at different times. We shall explain the detailed procedure through examples in \sects{cell}{wave}.

\subsection{Numerical implementation}
\label{numimp}

In order to adopt the above representation of condensation into our numerical framework, first recall from the discussion around \Eq{advdif} that we employ the split-step algorithm and first solve the advection--diffusion step to obtain the intermediate field $q_*$. This is then followed by solving the condensation step
\begin{equation}
\partialt{q_*} = -C.
\label{condstep}
\end{equation}
In the limit $\tau_c\rightarrow 0$, we can consistently set $\tau_c=\Delta t$ in the effective condensation \Eq{Ceff} where $\Delta t$ is the time step of the simulation. Assume $\Phio$ at the end of the advection--diffusion step is known and denote it by $\Phio(q'|x,y;t_{n+1})$. Discretizing \Eq{condstep} in time with $C$ given by \Eq{Ceff} leads to the condensation formula
\begin{multline}
q(x,y,t_{n+1}) = q_*(x,y,t_{n+1}) \\
- \int_{q_s(y)}^{\qmax}(q'-q_s)\Phio(q'|x,y;t_{n+1})\,\dd q'
\label{condeq}
\end{multline}
which gives the value of the specific humidity at the end of one full time step.

\begin{figure}
\centering
\includegraphics[width=0.31\textwidth]{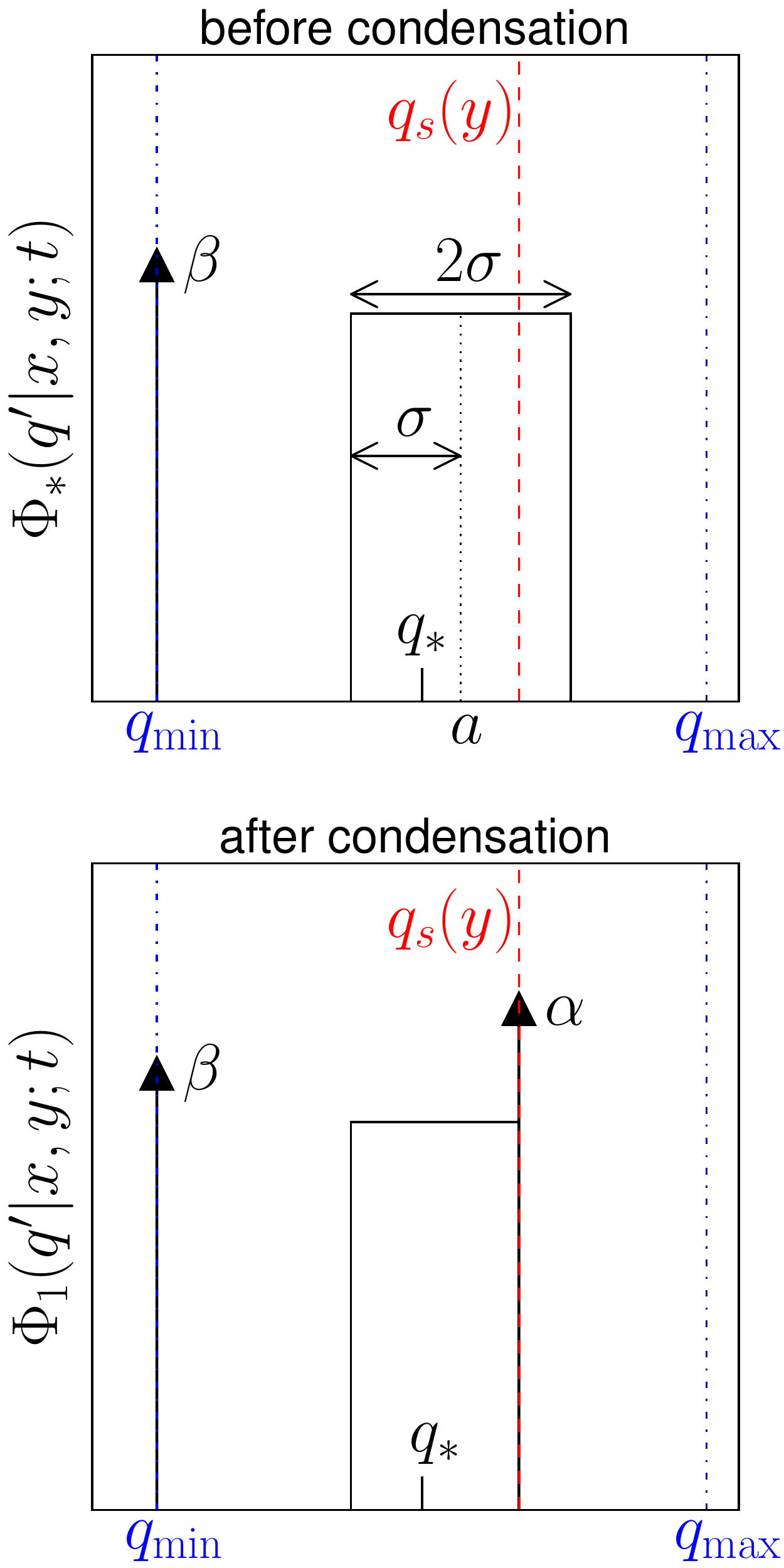}
\caption{Schematics of the condensation parameterization discussed in \sect{paracond}\ref{numimp}. The specific humidity $q_*$ at position (x,y) and time $t$ is thought of as the mean value of a distribution $\Phio$. The $\Phio$ illustrated here is defined in \Eq{phi0}. The action of rapid condensation collapses the part of $\Phio$ beyond the saturation limit $q_s$ onto a delta function at $q_s$.}
\label{condparab}
\end{figure}

Before we proceed further, we give a physical interpretation to \Eq{condeq} and also set the stage for specifying $\Phio$ in the next sections. The idea is to interpret the value of the specific humidity at a given grid point $(x,y)$ after the advection--diffusion step as the mean from an ensemble of parcels with specific humidity distribution $\Phio$, that is,
\begin{equation}
q_*(x,y,t_{n+1}) = \int_{\qmin}^{\qmax}q'\Phio(q'|x,y;t_{n+1})\,\dd q'.
\label{qstar}
\end{equation}
Note that some of these imagined parcels can have specific humidity higher than $q_s$ even if $q_* < q_s$. This is illustrated in the top panel of \Fig{condparab}. Next, we carry out rapid condensation ($\tau_c=0$)  on this ensemble to reduce the specific humidity of all supersaturated parcels to $q_s$. The distribution after condensation is:
\begin{equation}
\Phi_1(q'|x,y;t_{n+1}) =
\begin{cases}
\Phio(q'|x,y;t_{n+1}), & q' < q_s, \\
\alpha\delta(q-q_s),   & q' = q_s, \\
0,                     & q' > q_s,
\end{cases}
\label{phi1}
\end{equation}
where $\alpha$ is fixed by the normalization condition $\int_{\qmin}^{\qmax}\Phi_1\dd q'=1$. \FF{condparab} shows a schematic of this parameterized condensation. Finally, the specific humidity field at time $t_{n+1}$ is given by
\begin{align}
&q(x,y,t_{n+1}) = \int_{\qmin}^{\qmax}q'\Phi_1(q'|x,y;t_{n+1})\,\dd q' \nonumber \\
=\,&q_*(x,y,t_{n+1}) + \alpha q_s - \int_{q_s}^{\qmax}q'\Phio(q'|x,y;t_{n+1})\,\dd q'
\label{condeq1}
\end{align}
from which \Eq{condeq} follows.

\section{A steady overturning flow}
\label{cell}

In our first example of applying the condensation parameterization, we use the system introduced in \sect{ac} where moist air in a square cell is advected by the steady overturning flow $\vec u$ in \Eq{psi1}. The coarse-grained specific humidity field $q(x,y,t)$ in the Eulerian formulation is governed by the PDE (\ref{pde}). The fairly large diffusivity of $\kappa_q=10^{-1}$ magnifies the susceptibility to saturation in the coarse-grained model and puts the condensation parameterization to stringent test.

\begin{figure}
\centering
\includegraphics[width=0.45\textwidth]{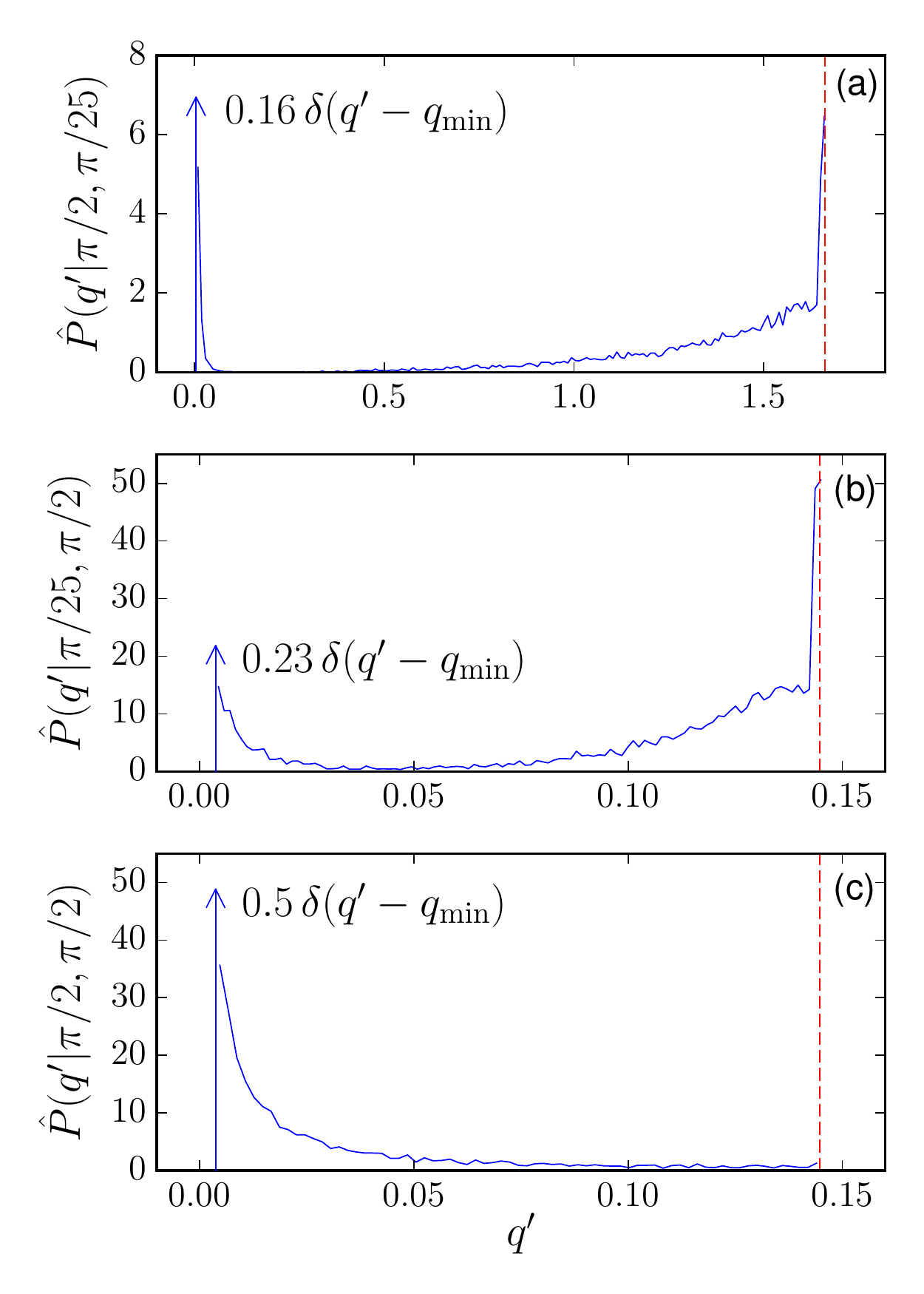}
\caption{Time-averaged probability distribution $\hat P(q'|x,y)$ of specific humidity at three different locations $(x,y)$ in the Monte Carlo simulation of \Fig{mcpde}(a). The solid line is the continuous component of the distribution and the arrow represents the discrete dry spike at $q'=\qmin$. The dashed line indicates the value of the local saturation value $q_s(x,y)$.}
\label{pdf_xy}
\end{figure}

Our first task is to choose an ansatz for $\Phio$. For this particular setup, subsidence of dry parcels from the upper part of the domain significantly affect the distribution of specific humidity. The driest parcels are created at the top boundary. They roam through the domain and maintain their dryness of $Q=\qmin$ until they hit the localized moisture source at the bottom boundary. As a consequence, we expect $\Phio$ to be composed of a dry spike of amplitude $\beta$ \citep{Sukhatme11,Tsang17} and a continuous part $\tPhio$:
\begin{equation}
\Phio(q'|x,y;t) = \beta(x,y,t)\delta(q-\qmin) + \tPhio(q'|x,y;t).
\label{phi0}
\end{equation}
In part for simplicity and in part because we expect the distribution of specific humidity to be smooth over a range of values (as we show later), we assume a top-hat shape for $\tPhio$ at all times. Referring to the top panel of \Fig{condparab}, $\tPhio$ is centered at $a$ with width $2\sigma$. Normalization condition gives:
\begin{equation}
\tPhio(q'|x,y;t) =
\begin{cases}
\!\displaystyle\frac{1-\beta}{2\sigma} \equiv h,  &  a-\sigma < q' < a + \sigma, \\[0.2cm]
0, & \text{otherwise}.
\end{cases}
\label{tphi0}
\end{equation}
Thus, $\Phio$ is fixed by the three parameters $(\beta,a,\sigma)$ which generally vary with both position and time. For comparison, \Fig{pdf_xy} shows the `true' (time-averaged) PDF, $\hat P(q'|x,y)$, from the Monte Carlo simulation of \Fig{mcpde}(a). To obtain the specific humidity field $q(x,y,t_{n+1})$ after condensation, we substitute \Eq{phi0} into the condensation formula \eqref{condeq}, or equivalently \eqref{condeq1}. Depending on the proportion of supersaturated parcels in the distribution, in other words, the location of $\tPhio$ relative to $q_s$, we have the three cases:
\begin{multline}
q(x,y,t_{n+1}) \\
= \begin{cases}
\beta\qmin  + (1-\beta)q_s
& \text{if } q_s \leqslant a-\sigma, \\[0.1cm]
\displaystyle
q_* - \frac{1-\beta}{4\sigma}(a+\sigma-q_s)^2
& \text{if } a-\sigma < q_s < a+\sigma, \\[0.1cm]
q_*
& \text{if } a+\sigma \leqslant q_s.
\end{cases}
\label{condtophat}
\end{multline}
We discuss how to determine $(\beta,a,\sigma)$ in the next sections with further technical details concerning some exceptional cases given in Appendix~A.

\subsection{Amplitude of the dry spike, $\beta$}

Because the dry parcels with $Q=\qmin$ simply move around the domain without undergoing condensation, it is particularly easy to calculate the amplitude of the dry spike $\beta$. Recalling from \Eq{phat} that $P=\pi^{-2}\hat P$, we substitute $P=\pi^{-2}\beta(x,y,t)\delta(q-\qmin)$ into \Eq{fptau}. Noting that $C(\qmin,y)=0$, we find $\beta$ satisfies
\begin{equation}
\partialt\beta + \vec u \cdot\nabla\beta = \kappa_b \nabla^2\beta.
\label{betaeq}
\end{equation}
Because of rapid condensation, parcels at the top boundary always have $Q=\qmin$. At the bottom boundary where the source is located, there is zero probability that $Q=\qmin$. Hence, the boundary conditions at the top and bottom are:
\begin{equation}
\beta(x,0,t) = 0, \quad \beta(x,\pi,t) = 1.
\end{equation}
At the other boundaries, the normal derivative vanishes. Unlike \Eq{qbareqn} for $\qbar$, \Eq{betaeq} is a closed equation in $\beta$ and can be solved to obtain $\beta$ for all $t$.

\subsection{Center of $\,\tPhio$, $a$}

The value of $a(x,y,t)$ is determined by the intermediate field $q_*(x,y,t)$ obtained at the end of the advection--diffusion stage in the split-step algorithm described around \Eq{advdif}. As mentioned before, $q_*$ is interpreted as the mean of the distribution $\Phio$. Substitution of \Eq{phi0} into \Eq{qstar} yields
\begin{equation}
a = \frac{q_*-\beta\qmin}{1-\beta}.
\label{aa}
\end{equation}
Note that no extra prognostic equation is introduced here.

\subsection{Width of $\,\tPhio$, $\sigma$}

The width of $\tPhio$ describes the subgrid-scale fluctuation of the specific humidity about its mean value before the action of condensation in each time step. To determine $\sigma(x,y,t)$, we use the second moment
\begin{equation}
\mu(x,y,t) = \int_{\qmin}^{\qmax} q'^2 \hat P(q'|x,y;t) \dd q'
\end{equation}
from the stochastic model. Consider the advection--diffusion of the ensemble of parcels without condensation from time $t_n$ to $t_{n+1}$ and assume the initial condition $\mu(x,y,t_n)$ is known. During this time, $\mu$ evolves to an intermediate value $\mu_*(x,y,t_{n+1})$ according to
\begin{equation}
\partialt\mu + \vec u \cdot\nabla\mu = \kappa_b \nabla^2\mu,
\label{mustar}
\end{equation}
which follows from \Eq{fptau}. The boundary conditions are
\begin{equation}
\mu(x,0,t) = \qmax^2
\end{equation}
and vanishing normal derivative at all other boundaries. Knowing $\mu_*$, we set the value of $\sigma$ in $\tPhio$ by requiring
\begin{equation}
\int_{\qmin}^{\qmax} q'^2 \Phio(q'|x,y;t_{n+1}) \dd q' = \mu_*(x,y,t_{n+1}).
\end{equation}
This gives
\begin{equation}
\sigma^2 = 3\left[\frac{\mu_*-\beta\qmin^2}{1-\beta} - \left(\frac{q_*-\beta\qmin}{1-\beta}\right)^{\!\!\!2}\,\right].
\label{sigma}
\end{equation}
After rapid condensation, the conditional PDF of specific humidity of the imagined ensemble becomes $\Phi_1$ given by \Eq{phi1} and depicted in the lower panel of \Fig{condparab}. Therefore, the initial condition for the next iteration is
\begin{align}
&\mu(x,y,t_{n+1}) = \int_{\qmin}^{\qmax} q'^2 \Phi_1(q'|x,y;t_{n+1}) \dd q' \nonumber \\[0.1cm]
=&
\begin{cases}
  \beta\qmin^2 + (1-\beta)q_s^2
& \text{if } q_s \leqslant a-\sigma, \\
  \mu_* + \alpha q_s^2 - \displaystyle \frac h3 [(a+\sigma)^3 - q_s^3]
& \text{if } a-\sigma < q_s < a + \sigma, \\
  \mu_*
& \text{if } a + \sigma \leqslant q_s
\end{cases}
\label{condmu}
\end{align}
with $\alpha$ defined in \Eq{phi1} and $h$ in \Eq{tphi0}.

\subsection{Results}

\begin{figure*}
\centering
\includegraphics[width=\textwidth]{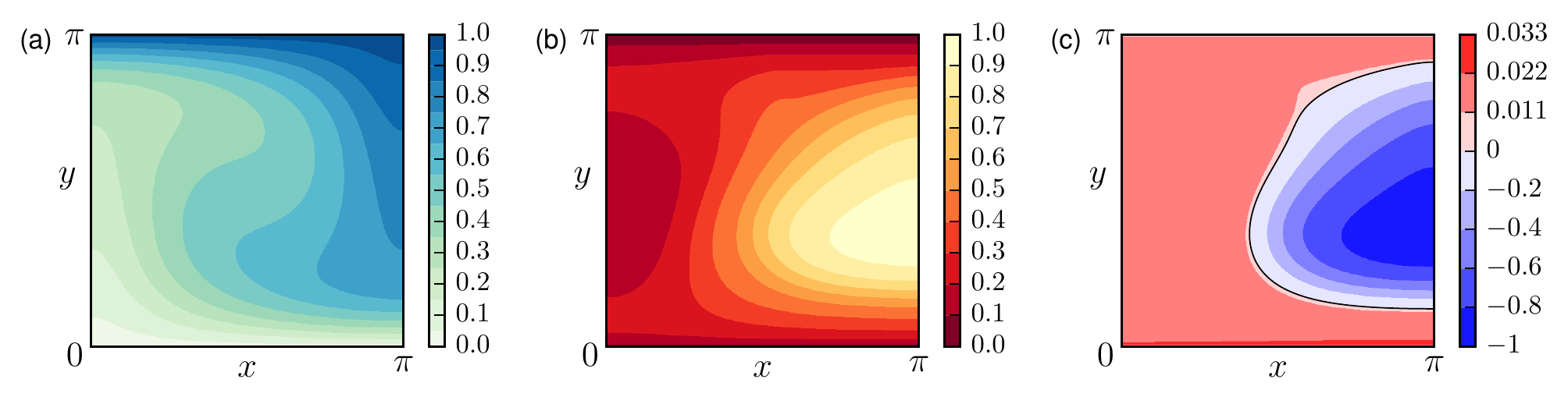}
\caption{Spatial structure of the parameters $(\beta,a,\sigma)$ that specify the assumed PDF $\Phio(q'|x,y;t)$ employed in the condensation parameterization in \sect{cell}. $\Phio$ is given by \Eqs{phi0}{tphi0} and illustrated in \Fig{condparab}. (a) The amplitude of the dry spike, $\beta$. (b) The distance of the center of $\tPhio$ from the local saturation limit (normalized by $q_s$), $(q_s-a)/q_s$. (c) Degree of supersaturation as measured by $(a+\sigma-q_s)/q_s$. The black solid line is $(a+\sigma-q_s)/q_s=0$ separating the unsaturated and the supersaturated regions.}
\label{abs}
\end{figure*}

Let us now summarize the full procedure. Given $q(t_n)$, $\beta(t_n)$ and $\mu(t_n)$ at time $t_n$ (with spatial arguments momentarily suppressed for clarity), we time-step forward the three advection--diffusion equations: (\ref{advdif}) for the moisture $q$ itself, (\ref{betaeq}) for the  amplitude of the dry spike $\beta$ and (\ref{mustar}) for the second moment $\mu$ of the `true' distribution $\hat P$. This gives $q_*(t_{n+1})$, $\beta(t_{n+1})$ and $\mu_*(t_{n+1})$ which in turn allows us to calculate $a$ and $\sigma$ from \eqref{aa} and \eqref{sigma}, respectively, and hence fully specify $\Phio$. Finally, the action of parameterized condensation depicted in \Fig{condparab} gives $q(t_{n+1})$ in \eqref{condtophat} and $\mu(t_{n+1})$ in \eqref{condmu}.  Note that the full Fokker--Planck equation is not solved (nor could it be). Rather, there are only as many evolution equations as there are parameters in the assumed PDF.

We run the parameterized system until it reaches the steady state. We first examine the spatial structure of the PDF parameters $(\beta,a,\sigma)$ in order to gain further insights into the parameterization process. \FF{abs}(a) plots the steady-state dry spike amplitude $\beta$. As expected from the boundary condition and the circulating flow pattern, $\beta\approx 1$ along the top and east edges while $\beta\ll 1$ along the bottom and west boundaries. A more surprising feature is that $\beta\approx 0.5$ for much of the area away from the boundaries. This means that in the central area, roughly half of the parcels in the imagined ensemble have the minimum specific humidity $\qmin$. This evinces the importance of subsidence of dry parcels by the random velocity. \FF{abs}(b) shows how close the center of $\tPhio$ is to the local saturation limit $q_s$ at different positions $(x,y)$. We find that $q_* < a < q_s$ for all $(x,y)$ with the first inequality follows directly from \Eq{aa}. \FF{abs}(c) plots the measure of supersaturation $(a+\sigma-q_s)/q_s$ of the ensemble. Recalling the schematic in \Fig{condparab}, we see that inside the red supersaturated region where $(a+\sigma-q_s)/q_s>0$, some of the imagined parcels are about $1-3\%$ over the local saturation limit. This is the region where the condensation parameterization is in action.

\begin{figure*}
\centering
\includegraphics[width=0.98\textwidth]{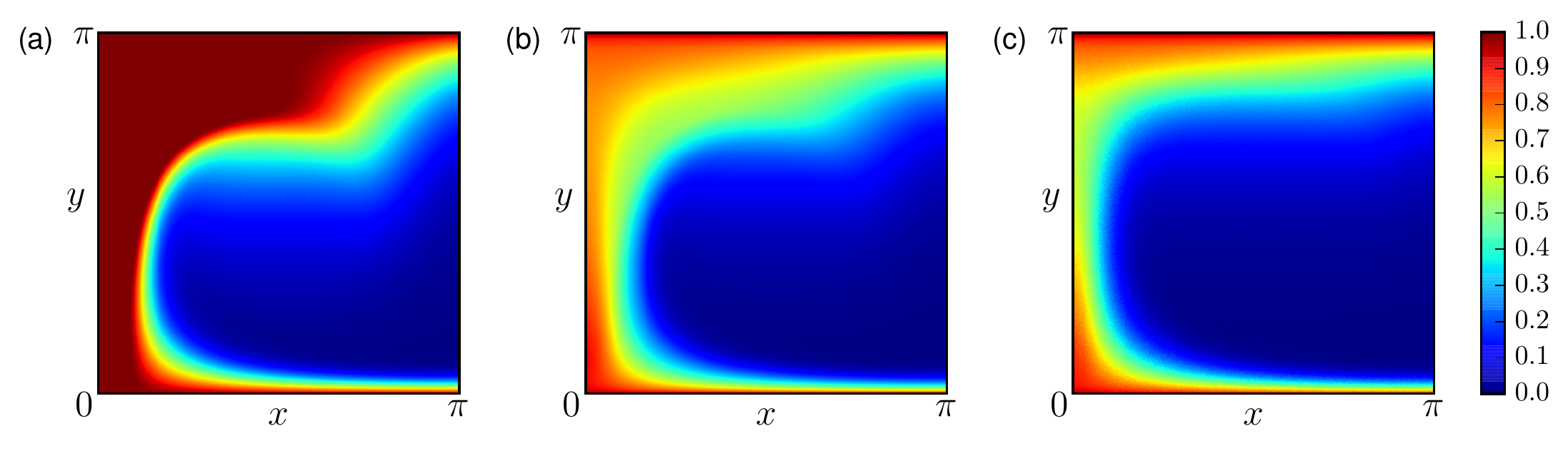}
\caption{Advection--condensation by the overturning flow \Eq{psi1} and with $\kappa=10^{-2}$. (a) Steady-state relative humidity field $r(x,y,t)$ at large $t$ from a solution of the Eulerian model \Eq{pde} without condensation parameterization. (b) Similar to (a) but with condensation parameterization implemented in the model. (c) Bin-averaged relative humidity field $\rhavg(x,y)$ derived from a Monte Carlo simulation of the Lagrangian model \Eq{sde} as described below \Eq{rhmc}.}
\label{kappa2}
\end{figure*}

We now assess the effectiveness of the condensation parameterization. \FF{mcpde}(d) shows the relative humidity field $\rhpara(x,y,t)$ of the parameterized system at a late time. In contrast to $r(x,y,t)$ from the unparameterized model with the same simulation parameters shown in \Fig{mcpde}(c), $\rhpara$ does not have large areas of complete saturation and approximately resembles the bin-averaged field $\rhavg(x,y)$ from the Lagrangian model in \Fig{mcpde}(b). The most visibly noticeable discrepancy appears inside the boundary layer near $x=0$. \FF{mcpde}(f) plots $\rhpara-\rhavg$ showing the biggest difference is located between such boundary layer and the central dry region. Comparing \Fig{mcpde}(f) to \Fig{mcpde}(e) and noting the difference in the color scales, we can see quantitatively the improvement due to the parameterization. \FF{kappa2} plots $r$, $\rhpara$ and $\rhavg$ for the case of small eddy diffusivity $\kappa=10^{-2}$ and shows the condensation parameterization is similarly effective.

\begin{figure*}
\centering
\includegraphics[width=\textwidth]{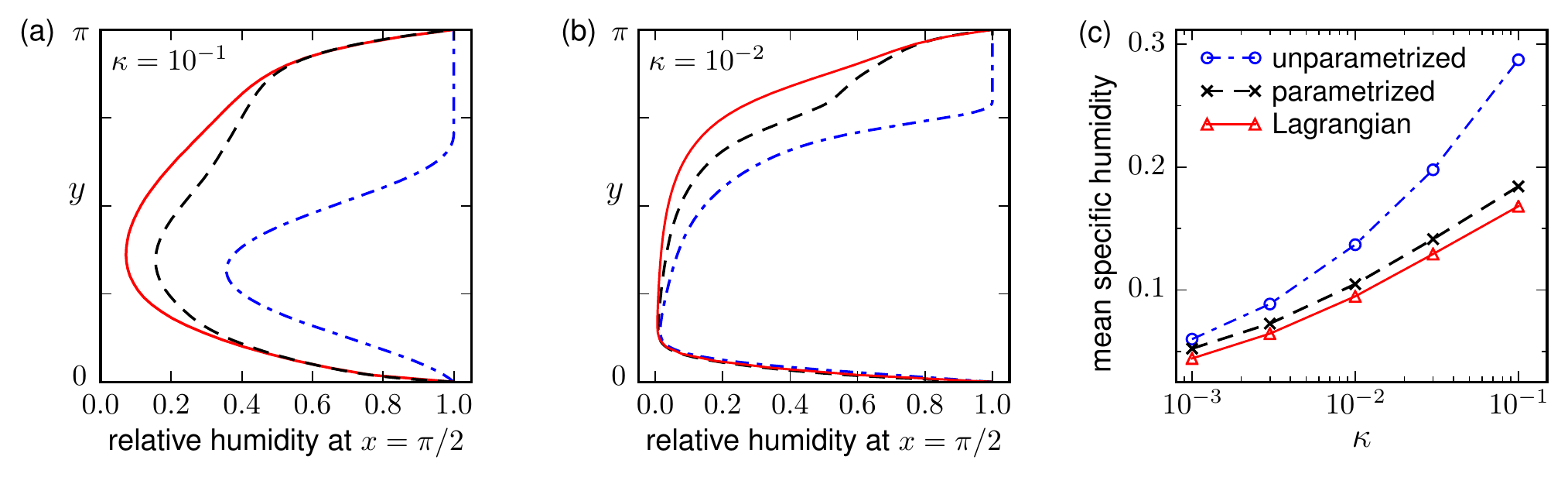}
\caption{Comparison of moisture content in the Eulerian model \Eq{pde}, with and without condensation parameterization, and in the Lagrangian model \Eq{sde} for the overturning cell of \sect{cell}. (a) Variation of relative humidity along $y$ at $x=\pi/2$ for $\kappa=10^{-1}$. (b) Same as (a) except for $\kappa=10^{-2}$. (c) Mean specific humidity, defined in \Eqs{meanq1}{meanq2}, for different $\kappa$.}
\label{rh_cell}
\end{figure*}

For further comparison, we plot the variation of the relative humidity along $y$ at a fixed $x=\pi/2$ for $\kappa=10^{-1}$ and $10^{-2}$ in \Fig{rh_cell}(a) and \ref{rh_cell}(b) respectively. For both values of $\kappa$, $\rhpara < r$ for all $y$. Near the top and bottom boundaries, $\rhpara$ and $\rhavg$ virtually have the same values whereas $\rhpara > \rhavg$ elsewhere. We also examine the total moisture content in the system by calculating the mean specific humidity. For the Lagrangian formulation, we have:
\begin{equation}
\text{mean specific humidity} = \avg{\frac1N\sum_{i=1}^N Q_i(t)}_{\!\!t}
\label{meanq1}
\end{equation}
where $N$ is the total number of parcels and $\avg{\cdot}_t$ indicates averaging over many snapshots in the statistically steady state. In the Eulerian formulation, using the steady solution at some large time $t_\infty$, we compute:
\begin{equation}
\text{mean specific humidity} = \frac{1}{\pi^2}\int_0^\pi\!\!\!\int_0^\pi\!q(x,y,t_\infty)\, \dd x \dd y.
\label{meanq2}
\end{equation}
\FF{rh_cell}(c) plots the mean specific humidity versus $\kappa$ for the three models studied here. The moisture content of the unparameterized Eulerian system is the highest and it also increases the fastest with $\kappa$. With the condensation parameterization implemented, the mean specific humidity in the Eulerian model and the rate at which it increases with $\kappa$ are both reduced to nearly the same as those in the Lagrangian model.

\begin{figure*}
\centering
\includegraphics[width=\textwidth]{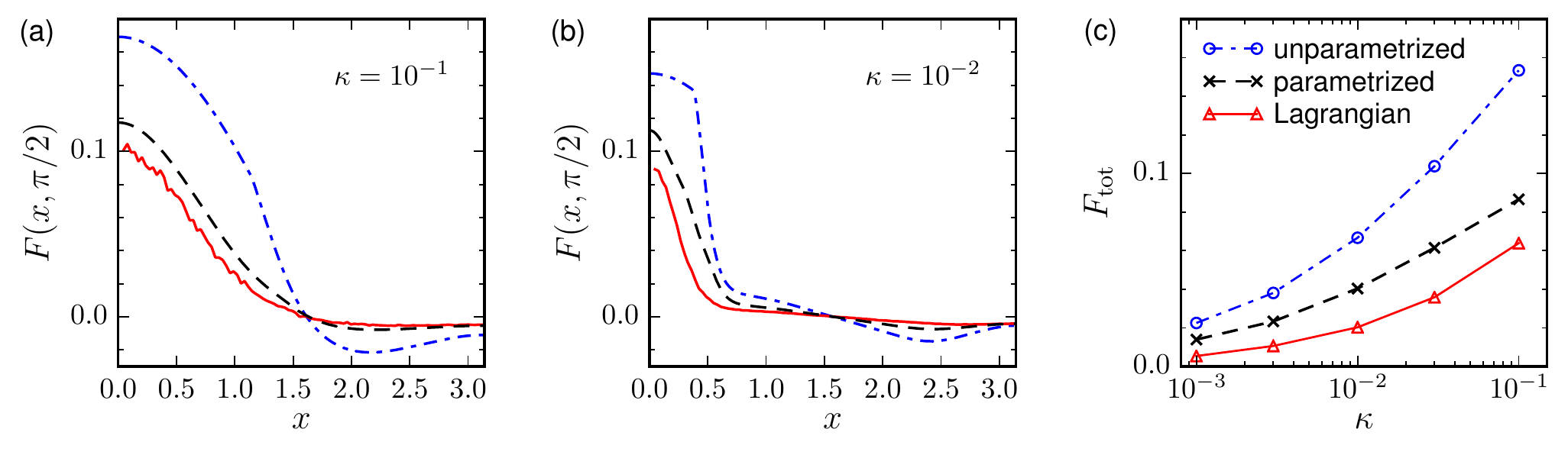}
\caption{Comparison of vertical moisture flux in the Eulerian model \Eq{pde}, with and without condensation parameterization, and in the Lagrangian model \Eq{sde} for the overturning cell of \sect{cell}. (a) Horizontal profile of the vertical moisture flux $F(x,\pi/2)$ across $y=\pi/2$ for $\kappa=10^{-1}$. (b) Same as (a) except for $\kappa=10^{-2}$. (c) Total vertical moisture flux $F_{\rm tot}$ across $y=\pi/2$, defined in \Eq{totalF}, for different $\kappa$.}
\label{flux_cell}
\end{figure*}

An important quantity in atmospheric moisture transport is the vertical moisture flux $F$. As the advecting velocity in our present system is steady, we focus on the steady-state flux. Hence, for the Eulerian formulation, we compute
\begin{equation}
F(x,y) = v(x,y) q(x,y,t_\infty) - \kappa_q \partialy{}q(x,y,t_\infty)
\end{equation}
where $t_\infty$ is some large time in the simulations. In the Lagrangian formulation, we estimate $F(x,y)$ by monitoring over a long period of time in the statistically steady state the specific humidity $Q$ of those parcels crossing a given altitude $y$. We relegate the detailed implementation of this diagnostic to Appendix B. \FF{flux_cell}(a) plots the horizontal profile of the vertical moisture flux $F(x,\pi/2)$ across $y=\pi/2$ for different models at $\kappa=10^{-1}$. \FF{flux_cell}(b) shows the same for the case of $\kappa=10^{-2}$. Generally, there is a large positive flux associated with the rising arm of the overturning cell for $x<\pi/2$ and a small negative flux in the descending arm for $x>\pi/2$. For both values of $\kappa$, the unparameterized system has the largest flux in magnitude $|F(x,\pi/2)|$ due to its high moisture content. When the condensation is parameterized in the Eulerian model, the magnitude of the flux is reduced and the profile $F(x,\pi/2)$ becomes close to that of the Lagrangian model. \FF{flux_cell}(c) plots the total vertical moisture flux across $y=\pi/2$,
\begin{equation}
F_{\rm tot} = \int_0^\pi F(x,\pi/2)\, \dd x,
\vspace{-0.04cm}
\label{totalF}
\end{equation}
for different $\kappa$. The total flux generally increases with $\kappa$. Not surprisingly, the unparameterized Eulerian model produces the largest $F_{\rm tot}$ at all $\kappa$. With the condensation parameterized, $F_{\rm tot}$ in the Eulerian model is reduced by about $50\%$. The smallest $F_{\rm tot}$ is observed in the Lagrangian model.

\section{An unsteady channel flow}
\label{wave}

\begin{figure*}
\centering
\includegraphics[width=0.98\textwidth]{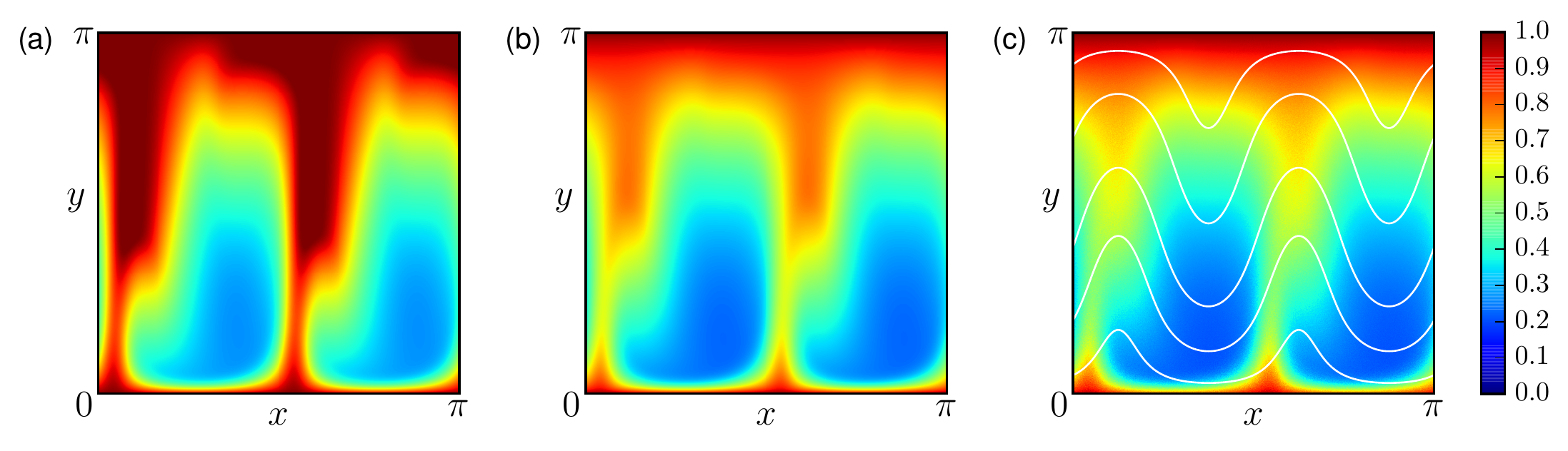}
\caption{Relative humidity field at time $t=14.5$ for the unsteady channel flow of \sect{wave} with $\kappa=10^{-1}$. (a) $r(x,y,t)$ from the Eulerian model \Eq{pde} with no parameterization; (b) $\rhpara(x,y,t)$ from the Eulerian model with condensation parameterization; (c) bin-averaged field $\rhavg(x,y,t)$ obtained from a Monte Carlo simulation of the Lagrangian model \Eq{sde}. The solid lines are streamlines of \Eq{psichnl} at $t=14.5$.}
\label{rhchnl}
\end{figure*}

For our second example, we apply the condensation parameterization to a time-dependent flow. We introduce a configuration that roughly mimics the transport of moisture by baroclinic eddies along moist isentropic surfaces in mid-latitudes \citep{Vallis17}. Consider a channel of width $\pi$ in the $y$-direction and periodic in the $x$-direction. The streamfunction of the unsteady velocity $(u,v)=(-\partial_y\psi,\partial_x\psi)$ in the channel is taken to be
\begin{equation}
\psi(x,y,t) = -U y + \Psi(t) \sin(k x - \omega t) \sin l y
\label{psichnl}
\end{equation}
where
\begin{equation}
\Psi(t) = \Psi_0[1 - \delta \cos(\gamma\omega t)].
\end{equation}
The values of the parameters are: $U=2\pi$, $k=4$, $l=1$, $\omega=4\pi$, $\Psi_0=3\pi/2$, $\delta=0.5$ and $\gamma=0.75$. We choose $U/\Psi_0>1$ to ensure all streamlines are open and wrap around the periodic $x$-direction. \FF{rhchnl}(c) shows several streamlines of \Eq{psichnl} at one instance of time. The waviness of the streamlines, controlled by $\Psi(t)$, varies with time as the whole pattern propagates eastward. We use the saturation profile $q_s$ given in \Eq{qsat} with $y$ interpreted as the meridional direction. Hence, we take $T_{\max}=20\,^{\circ}$C and $T_{\min}=-10\,^{\circ}$C which gives $\qmax=1.39$ and $\qmin=0.17$. We once again assume an evaporation source that saturates air parcels is located along $y=0$. At $y=\pi$, we have $\partial_y q=0$. The domain is initially saturated. Advected by the time-periodic velocity, the moisture field eventually reaches a time-periodic state that varies at the same frequency $f_0=0.5$ as the velocity.

We first consider the Lagrangian formulation of the problem. As in the previous example, we perform Monte Carlo simulation of the stochastic system (\ref{sde}) and calculate the bin-averaged relative humidity field $\rhavg(x,y,t)$ from the data. \FF{rhchnl}(c) shows $\rhavg$ for $\kappa_b=10^{-1}$ at a late time after the transient, specifically $t=14.5$. We see that the areas along the top and bottom edges are close to saturation. Large regions of low relative humidity are formed in the middle of the channel. These dry patches are separated by tongues of humid air erupting periodically from the top and bottom boundary layers as the general large-scale pattern propagates eastward. Interestingly, the jets of humid air emerging from the bottom boundary are filamentous, creating sharp gradients in humidity. Similar features have been reported in more complex models of moisture decay on isentropic surfaces \citep{Yang94}.

\begin{figure}
\centering
\includegraphics[width=0.43\textwidth]{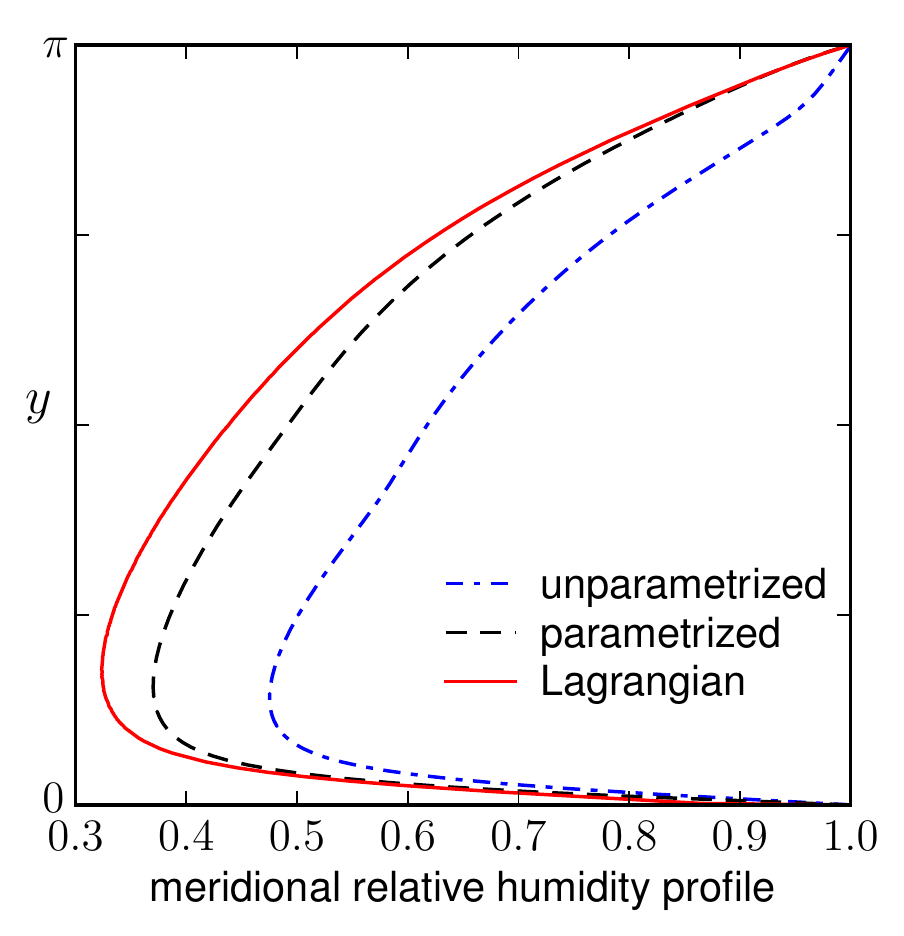}
\caption{Meridional profile of relative humidity for the channel flow in \sect{wave}, obtained by averaging the relative humidity field over the zonal direction $x$ and time $t$ after the initial transient.}
\label{rhy_chnl}
\end{figure}

\begin{figure}
\centering
\includegraphics[width=0.47\textwidth]{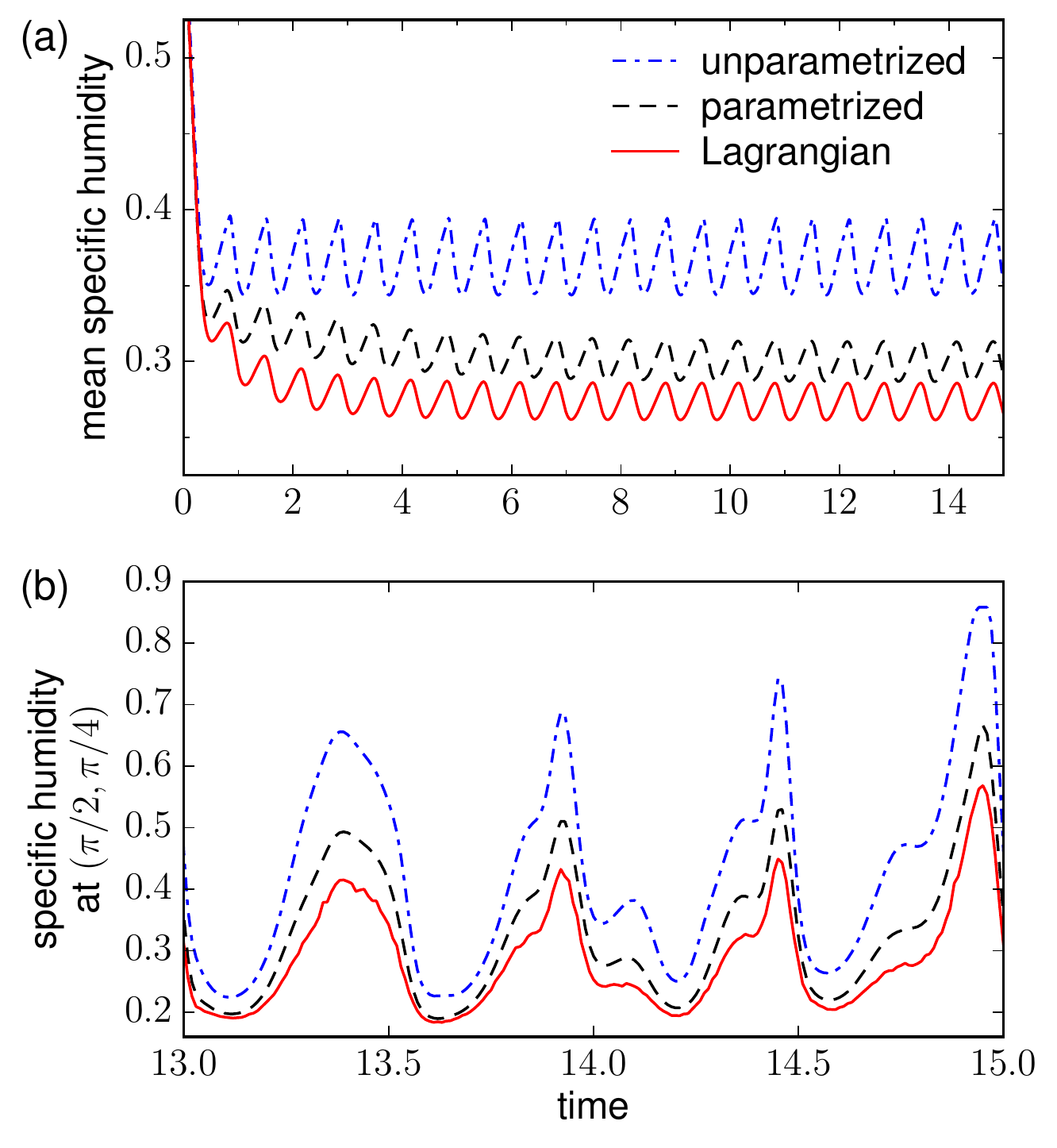}
\caption{For the channel flow in \sect{wave}: (a) Time evolution of the mean specific humidity defined in \Eqs{meanq1}{meanq2}. The period of variation $f_0^{-1}=2$ is the same as that of the advecting velocity \Eq{psichnl}. There are three peaks in each period. (b) Time variation of the specific humidity at the location $(x,y)=(\pi/2,\pi/4)$ over one period.}
\label{qavg_chnl}
\end{figure}

Turning to the Eulerian formulation, we recall that the system is now governed by the PDE (\ref{pde}). \FF{rhchnl}(a) shows the relative humidity field $r(x,y,t)$ obtained from a solution of \eqref{pde} with unparameterized rapid condensation and \Fig{rhchnl}(b) shows $\rhpara(x,y,t)$ for the case when condensation is parameterized. We once again use the ansatz \Eq{phi0} in our parameterization with the three parameters determined by the same procedure described in the previous section. In both figures, $\kappa_q=10^{-1}$ and $t=14.5$, i.e. the same diffusivity and time instance as in \Fig{rhchnl}(c). All three relative humidity fields in \Fig{rhchnl} display the same general structure of high and low values. However, large areas of complete saturation can be seen in the unparameterized Eulerian model. Furthermore, the minimum relative humidity inside the dry patches is about $20-30\%$ higher than those in the other two models. \FF{rhy_chnl} shows the meridional relative humidity profile obtained by averaging over the zonal direction $x$ and time $t$ after the initial transient. The difference in the magnitude of the relative humidity minimum in the profile from the three models is obvious. The high moisture content in the unparameterized system is also evident in \Fig{qavg_chnl}(a) which plots the time evolution of the mean specific humidity. Interestingly, \Fig{qavg_chnl}(a) also shows that the mean specific humidity oscillates with a much larger amplitude in the unparameterized system. \FF{qavg_chnl}(b) shows how the specific humidity changes with time at one particular location $(x,y)=(\pi/2,\pi/4)$ over a single period of variation. We see that the evolution from the parameterized Eulerian model approximately follows the one from the Lagrangian model. This demonstrates that these two systems are not only close to each other in the average sense but actually have similar spatio-temporal behavior.

\section{Parameterization in atmospheric models}
\label{sec:slmpara}

Probabilistic (or statistical) schemes are often used in atmospheric general circulation models (GCMs), and sometimes cloud resolving models, to parameterize subgrid-scale moisture variability. As discussed in the Introduction, these schemes often employ turbulence closures to obtain the moments required to fix the assumed PDF. In view of our results that an Eulerian model with probabilistic condensation, namely \eqref{Ceff}, can successfully mimic a Lagrangian model, we suggest a strategy that makes use of a stochastic Lagrangian model instead of turbulence closures. Of course, at a fundamental level the two methodologies are not so different for there is a close relationship between stochastic Lagrangian models and turbulence models, in particular second-moment closures \citep{Pope94b}. However, our method avoids the `intermediate' step of constructing a closure. Stochastic Lagrangian models are also often used as models of turbulent diffusion \citep{Rodean96} and the dispersion of passive, non-reactive scalars in the atmosphere \citep{Wilson96}.  It may also be noted \citep[e.g.,][]{Pope94} that the Lagrangian framework is especially fit for modeling reactive flows, and condensation can be considered mathematically as a form of reaction. 

To construct a parameterization that might be used in an atmospheric GCM, one would first construct a stochastic Lagrangian model of water vapor transport, such as \eqref{sde}, for the atmospheric flow under consideration. Imagine an ensemble of moist parcels advected by the flow. Each parcel carries a set of thermodynamical variables (e.g., specific humidity and potential temperature) that evolves due to moist processes such as condensation and evaporation. The parcel moves with velocity $\vec V + \vec V'$. The large-scale velocity $\vec V$ is interpolated from the Eulerian velocity field $\vec u$ provided by the atmospheric model. To complete the model, we assume the salient properties of the small-scale parcel velocity $\vec V'$ can be captured by a suitably chosen random process---so that our stochastic system is a good representation of the moisture dynamics. This is a non-trivial issue and the details will depend on the particular application. For example, in some cases, the Markovian, i.e. memoryless, Brownian process is a sufficiently good model while in other cases, it may be necessary to consider time-correlated or non-Markovian random processes. 

If computational cost were not a constraint, we could perform Monte Carlo simulation (as in \Fig{mcpde}(b)) or solve for the governing PDF $P$ using the Fokker--Planck equation of the stochastic system. The mean humidity field can then be computed. However, the large number of parcels required to obtain good statistics and the high dimension of the Fokker--Planck equation render these propositions impractical (and, in any case, were computational costs not a consideration one could perform extremely high resolution Eulerian simulations without the need to parameterize subgrid-scale motion). Instead, we apply the effective condensation $\Ceff$, given in \Eq{Ceff}, to the evolution PDE for the Eulerian humidity field $q$ in the atmospheric model, e.g.
\begin{equation}
\partialt{q} + \vec u\cdot\nabla q  = \nabla\cdot(D \nabla q) + S - \Ceff.
\label{parapde}
\end{equation}
At each time step, the parameters of the assumed PDF $\Phio$ embedded in $\Ceff$ are determined, as described in \sect{cell}, by matching a certain number of moments of $\Phio$ to those of $P$. Investigation in \sects{cell}{wave} suggests that \Eq{parapde} will produce similar results to the stochastic Lagrangian model. Therefore we can use \Eq{parapde} in place of the Lagrangian model to parameterize the actual moisture transport.

In the procedure presented here, the stochastic Lagrangian model forms the foundation of an integrated parameterization scheme. It provides the theoretical basis for the effective condensation \eqref{Ceff} (as discussed in \sect{paracond}) as well as fixing the parameters in $\Phio$. It is in the explicit use of an underlying Lagrangian model that our method differs from other parameterization schemes in which moisture variability is not incorporated so directly. A possible advantage of the approach is the flexibility to incorporate different Lagrangian dynamics into the parameterization through the stochastic model \citep{Wilson96,Sawford01}. Results from atmospheric tracer experiments \citep{Stohl98} or novel theoretical transport models such as anomalous fractional diffusion \citep{Goulart17} may also be adopted into the scheme. 

The potential disadvantage of the scheme is that it requires extra prognostic equations in addition to the one for the humidity $q$. Generally, the total number of equations equals the number of undetermined parameters in the assumed PDF, thus in our examples where $\Phio$ has three parameters, two additional equations are introduced, namely \Eqs{betaeq}{mustar}. However,these prognostic equations are solved at the same resolution as the other variables in the atmospheric model. Modern GCMs often have a very large number of prognostic equations, especially if the model has an aerosol scheme, so that the additional expense of our scheme would be relatively small.

\section{Summary and conclusions}
\label{sec:conclude}

The representation of subgrid-scale condensation of moisture in climate or weather models is a matter of both theoretical interest and considerable practical concern. Using the simple advection--condensation model, \eqref{pde}, we have shown that, without any condensation parameterization, a coarse-grained PDE model tends to retain excessive moisture and develop large regions of high humidity. Fundamentally, this is because the nonlinear condensation process and the coarse-graining operation do not commute and local fluctuations are therefore lost when moisture is represented by a coarse-grained field, as illustrated in \Fig{condfig}. On the other hand, the comparison in \Fig{mcpde} shows that a Lagrangian formulation, where air parcels tagged with a humidity variable are tracked, is able to account for small-scale fluctuations, as found in nature.

It is, however, possible for an Eulerian model to produce results similar to the corresponding Lagrangian model if subgrid-scale moisture variability is properly introduced. \sect{paracond} presents a way of achieving this using a probabilistic condensation parameterization given in \eqref{Ceff}. This mimics the Lagrangian representation of condensation in \eqref{qbareqn} by using an assumed PDF of humidity, with the parameters of the PDF being given through the use of the Fokker--Planck equation that governs the stochastic Lagrangian model. In both the simple single-celled circulation patterns shown in \Fig{kappa2} and in the unsteady channel flow shown in \Fig{rhchnl}, we see that this methodology reduces the excessive saturation in the Eulerian model, allowing it to produce moisture distribution close to that of the original Lagrangian model, obtained by a Monte Carlo simulation that explicitly follows the moist parcels. That the Eulerian model with a probabilistic parameterization can mimic the explicit Lagrangian model is a quite stringent test for the method. 

The use of such a parameterization of condensation in a GCM trying to model real atmospheric flows would be rather more complex than our examples, but would follow the same methodology. That is, presuming that trajectories in the atmosphere can be modeled by a stochastic Lagrangian model, a coarse-resolution Eulerian model which incorporates subgrid-scale Lagrangian information via the probabilistic condensation \eqref{Ceff} could be used to parameterize water vapor transport efficiently, as discussed in \sect{sec:slmpara}. The first step is to construct a stochastic model, analogous to \eqref{sde} but with a more complete  thermodynamics and, potentially, non-Markovian dynamics. The second step is to choose an ansatz for the PDF of the thermodynamic variables, a PDF that is determined by a small number of parameters, and to use that ansatz in the Fokker--Planck equation to determine those parameters. In the examples we computed, we chose the PDF to be a dry spike plus a continuous (top hat) component with a finite width, but other choices, with more free parameters, are possible.  Since explicitly computing the Lagrangian model with a Monte Carlo simulation will not generally be possible in such cases the efficacy of the choices will ultimately be determined by comparison with observation. 

Testing this method in a range of models of varying complexity is the next step, starting from fairly idealized settings such as non-precipitating moist Rayleigh--B\'enard convection \citep{Pauluis13} or a minimal precipitating convection model \citep{Duenas13}.  Note too that the general idea behind the method is not limited to the condensation `reaction' --- its applicability to the parameterization of mixdown time in atmospheric chemical transport \citep{Thuburn97} could also be explored.

\acknowledgments
YKT is indebted to Steven B\"oing for many helpful discussions. This work is supported by a Feasibility Grant from the EPSRC network Research on Changes of Variability and Environmental Risk (ReCoVER). YKT also thanks the EPSRC-funded Past Earth Network (Grant number EP/M008363/1) for hosting a Writing Retreat during which part of this paper is written. GKV also acknowledges support from the Royal Society (Wolfson Foundation), the Leverhulme Trust, and NERC.

\appendix[A]
\appendixtitle{Exceptional cases in the determination of $(\beta,a,\sigma)$}

As discussed in \sects{cell}{wave}, we take the Lagrangian model \Eq{sde} as a good parameterization of the moist dynamics in the two systems considered there. We then determine the parameters $(\beta,a,\sigma)$ of the assumed-shaped PDF $\Phio(q'|x,y,t)$ in the effective condensation \Eq{Ceff} using information from the Lagrangian model. Although it rarely occurs in practice, two issues could in principle arise because $\Phio$ generally does not satisfy the Fokker--Planck equation of the Lagrangian model. First, we would have $\sigma^2 < 0$ in \Eq{sigma} if
\begin{equation}
\mu_* > q_*^2 + \frac{\beta}{1-\beta}(q_*-\qmin)^2.
\end{equation}
If this occurs, we set $\sigma=0$. Second, when $q_*$, and hence $a$, gets close to $\qmin$ or $\qmax$, it is possible for a portion of $\tPhio$ (the continuous top-hat component of $\Phio$) to lie outside the range $[\qmin,\qmax]$. When this happens, we reduce $\sigma$ so that either $a-\sigma=\qmin$ or $a+\sigma=\qmax$. In the highly unlikely case that $a>\qmax$, we set $\sigma=0$ and adjust $\beta$ to make $a=\qmax$.

\appendix[B]
\appendixtitle{Estimation of vertical moisture flux in Monte Carlo simulations}

Consider a Monte Carlo simulation using $N$ parcels in a $\pi\times\pi$ domain. Following \cite{Tsang17}, we estimate the vertical moisture flux $F(x,y,t)$ as follow. Assume between time $t$ and $t+\Delta t$, there are $N_p(x,y,t)$ parcels crossing a given height $y$ in either direction and whose $x$-positions $X_i(t)$ lies between $[x-\Delta x/2,x+\Delta x/2]$. Let $\xi_i(t)$ be the sign of $\dd Y_i/\dd t$, then
\begin{gather}
F(x,y,t)
=\frac{\pi^2}{N\Delta x\Delta t} \sum_{i=1}^{N_p} \xi_i\, Q^\dagger_i(t) \nonumber \\
\quad\text{where}\quad
Q^\dagger_i(t) = 
\begin{cases}
\min[Q_i(t),q_s(y)] & \text{if } \xi_i >0, \\
Q_i(t) & \text{if } \xi_i < 0.
\end{cases}
\end{gather}
The statistically steady $F(x,y)$ is then obtained by averaging $F(x,y,t)$ over $t$.

\balance
\bibliographystyle{ametsoc2014}
\bibliography{condense_para}

\end{document}